\documentstyle[12pt]{article}
\newcommand {\beq}{\begin{equation}}
\newcommand {\eeq}{\end{equation}}
\newcommand {\beqa}{\begin{eqnarray}}
\newcommand {\eeqa}{\end{eqnarray}}
\newcommand {\n}{\nonumber \\}

\begin{document}
\setlength{\oddsidemargin}{0cm}
\setlength{\baselineskip}{7mm}

\begin{titlepage}
 \renewcommand{\thefootnote}{\fnsymbol{footnote}}
$\mbox{ }$
\begin{flushright}
\begin{tabular}{l}
hep-th/9908141\\
KEK-TH-637 \\
KUNS-1595\\
\end{tabular}
\end{flushright}

~~\\
~~\\
~~\\

\vspace*{0cm}
    \begin{Large}
       \vspace{2cm}
       \begin{center}
         {Noncommutative Yang-Mills in IIB Matrix Model}      \\
       \end{center}
    \end{Large}

  \vspace{1cm}

\begin{center}
           Hajime A{\sc oki}$^{1)}$\footnote
           {
e-mail address : haoki@ccthmail.kek.jp},
           Nobuyuki I{\sc shibashi}$^{1)}$\footnote
           {
e-mail address : ishibashi@post.kek.jp}
           Satoshi I{\sc so}$^{1)}$\footnote
           {
e-mail address : satoshi.iso@kek.jp},\\
           Hikaru K{\sc awai}$^{2)}$\footnote
           {
e-mail address : hkawai@gauge.scphys.kyoto-u.ac.jp},
           Yoshihisa K{\sc itazawa}$^{1)}$\footnote
           {
e-mail address : kitazawa@post.kek.jp}{\sc and}
           Tsukasa T{\sc ada}$^{3)}$\footnote
           {
e-mail address : tada@tohwa-u.ac.jp}\\
        $^{1)}$ {\it High Energy Accelerator Research Organization (KEK),}\\
               {\it Tsukuba, Ibaraki 305-0801, Japan} \\
        $^{2)}$ {\it Department of Physics, Kyoto University,
Kyoto 606-8502, Japan}\\
        $^{3)}$ {\it Tohwa Institute for Science, Tohwa University,
Fukuoka 815-8510, Japan}\\
\end{center}

\vfill

\begin{abstract}
\noindent
\end{abstract}
We show that twisted reduced models can be interpreted as
noncommutative Yang-Mills theory.
Based upon this correspondence, we obtain noncommutative Yang-Mills theory
with D-brane backgrounds in IIB matrix model.
We propose that IIB matrix model with D-brane backgrounds
serve as a concrete definition of noncommutative Yang-Mills.
We investigate D-instanton solutions as local excitations on D3-branes.
When instantons overlap, their interaction can be well described
in gauge theory and $AdS$/CFT correspondence.
We show that IIB matrix model gives us the consistent potential with
IIB supergravity when they are well separated.

\vfill
\end{titlepage}
\vfil\eject

\section{Introduction}
\setcounter{equation}{0}
A large $N$ reduced model has been proposed as a nonperturbative
formulation of type IIB superstring theory\cite{IKKT}\cite{FKKT}.
It is defined by the following action:
\beq
S  =  -{1\over g^2}Tr({1\over 4}[A_{\mu},A_{\nu}][A^{\mu},A^{\nu}]
+{1\over 2}\bar{\psi}\Gamma ^{\mu}[A_{\mu},\psi ]) .
\label{action}
\eeq
It is a large $N$ reduced model of ten dimensional super Yang-Mills
theory.
Here $\psi$ is a ten dimensional Majorana-Weyl spinor field, and
$A_{\mu}$ and $\psi$ are $N \times N$ Hermitian matrices.
It is formulated in a manifestly covariant way which enables us
to study the nonperturbative issues of superstring theory.
In fact we can in principle predict the dimensionality of spacetime,
the gauge group and the matter contents by solving this model.
We have already initiated such investigations
in \cite{AIKKT}\cite{ISK}.
We refer our recent review for more
detailed expositions and references\cite{review}.
We also note a deep connection between our approach and
noncommutative geometry\cite{Connes}\cite{CDS}.

This action can be related to the Green-Schwarz action of
superstring\cite{GS}
by using the semiclassical correspondence in the large $N$ limit:
\beqa
-i[\;,\;] &\rightarrow&  \{\;,\;\}, \n
Tr &\rightarrow& \int {d^2 \sigma }\sqrt{\hat{g}} .
\label{correspondence}
\eeqa
In fact eq.(\ref{action}) reduces to the Green-Schwarz action
in the Schild gauge\cite{Schild}\cite{Zacos}\cite{Bars}:
\beq
S_{Schild}=\int d^2\sigma [\sqrt{\hat{g}}\alpha(
\frac{1}{4}\{X^{\mu},X^{\nu}\}^2
-\frac{i}{2}\bar{\psi}\Gamma^{\mu}\{X^{\mu},\psi\})
+\beta \sqrt{\hat{g}}].
\label{SSchild}
\eeq
Through this correspondence, the eigenvalues of $A_{\mu}$ matrices are
identified with the spacetime coordinates $X^{\mu}(\sigma )$.
The $\cal{N}$=2 supersymmetry manifests itself in  $S_{Schild}$ as
\beqa
\delta^{(1)}\psi &=& -\frac{1}{2}
                      \sigma_{\mu\nu}\Gamma^{\mu\nu}\epsilon  ,\n
\delta^{(1)} X^{\mu} &=& i\bar{\epsilon }\Gamma^{\mu}\psi ,
\label{SSchildsym1}
\eeqa
and
\beqa
\delta^{(2)}\psi &=& \xi ,\n
\delta^{(2)} X^{\mu} &=& 0 .
\label{SSchildsym2}
\eeqa
The $\cal{N}$=2 supersymmetry (\ref{SSchildsym1}) and (\ref{SSchildsym2})
is directly translated into
the symmetry of $S$ as
\beqa
\delta^{(1)}\psi &=& \frac{i}{2}
                     [A_{\mu},A_{\nu}]\Gamma^{\mu\nu}\epsilon ,\n
\delta^{(1)} A_{\mu} &=& i\bar{\epsilon }\Gamma^{\mu}\psi ,
\label{Ssym1}
\eeqa
and
\beqa
\delta^{(2)}\psi &=& \xi ,\n
\delta^{(2)} A_{\mu} &=& 0.
\label{Ssym2}
\eeqa

If we take a linear combination of $\delta^{(1)}$ and $\delta^{(2)}$ as
\beqa
\tilde{\delta}^{(1)}&=&\delta^{(1)}+\delta^{(2)}, \n
\tilde{\delta}^{(2)}&=&i(\delta^{(1)}-\delta^{(2)}),
\eeqa
we obtain the $\cal{N}$=2 supersymmetry algebra.
\beqa
(\tilde{\delta}^{(i)}_{\epsilon}\tilde{\delta}^{(j)}_{\xi}
    -\tilde{\delta}^{(j)}_{\xi}\tilde{\delta}^{(i)}_{\epsilon})\psi   &=&0 ,\n
(\tilde{\delta}^{(i)}_{\epsilon}\tilde{\delta}^{(j)}_{\xi}
    -\tilde{\delta}^{(j)}_{\xi}\tilde{\delta}^{(i)}_{\epsilon})A_{\mu}&=&
                                 2i\bar{\epsilon}\Gamma^{\mu}\xi
\delta_{ij}.
\eeqa
The $\cal{N}$=2 supersymmetry is a crucial element of superstring theory.
It imposes strong constraints on the spectra of particles.
Furthermore it determines the structure of the interactions uniquely in
the light-cone string field theory\cite{FKKT}.
The IIB matrix model is a nonperturbative formulation which possesses
such a symmetry. Therefore it has a very good chance to capture the
universality class of IIB superstring theory.
These symmetry considerations force us to interpret
the eigenvalues of $A_{\mu}$ as the space-time coordinates.
Note that our argument is independent of the D-brane interpretations
which are inevitably of semiclassical nature\cite{Polchinski}.

We recall the typical classical solutions of (\ref{action})
which represent infinitely long static D-strings.
When $\psi=0$, the equation of motion of (\ref{SSchild}) is
\beq
\{X^{\mu},\{X^{\mu},X^{\nu}\}\}=0.
\label{SSchildEOM}
\eeq
Corresponding to this,
the equation of motion of (\ref{action}) is
\beq
[A_{\mu},[A_{\mu},A_{\nu}]]=0.
\label{SEOM}
\eeq
We can easily construct a solution of (\ref{SSchildEOM}), which represents
a static D-string extending straight in the $X^1$ direction:
\beqa
X^0&=&T \tau ,\n
X^1&=&\frac{L}{2\pi}\sigma ,\n
\mbox{other }X^{\mu} \mbox{'s}&=& 0,
\label{Schildstaticsolution}
\eeqa
where $T$ and $L$ are large enough  extensions of a D-string and
\beqa
0 \leq &\tau& \leq 1, \n
0 \leq &\sigma& \leq 2\pi.
\eeqa
Considering the relation between the commutator and the Poisson bracket,
we obtain a solution of (\ref{SEOM}) corresponding to the above one as
follows:
\beqa
A_0 &=& \frac{T}{\sqrt{2\pi n}}\hat{q} \equiv \hat{p}_0 ,\n
A_1 &=& \frac{L}{\sqrt{2\pi n}}\hat{p} \equiv \hat{p}_1 ,\n
\mbox{other }A_{\mu} \mbox{'s}&=& 0,
\label{Sstaticsolution}
\eeqa
where $T$ and $L$ are large enough extensions of a D-string, and $\hat{q}$
and $\hat{p}$ are
$n \times n$ Hermitian matrices having the following commutation relation
and the eigenvalue distributions:
\beq
[\hat{q},\hat{p}]=i,\\
\label{qpcommutator}
\eeq
and
\beqa
&~&0 \leq \hat{q} \leq {\sqrt{2\pi n}},\n
&~&0 \leq \hat{p} \leq {\sqrt{2\pi n}}.
\label{qp}
\eeqa
Strictly speaking such $\hat{p}$ and $\hat{q}$ do not exist for finite
values of $n$.
For large values of $n$, however, we expect that (\ref{qpcommutator}) can
be approximately satisfied, because it is nothing but the canonical
commutation relation. As is well-known in the correspondence between the
classical and quantum mechanics, the total area of the $p-q$ phase space
is equal to $2\pi$
multiplied by the dimension of the representation. In this sense (\ref{qp})
indicates that $\hat{p}$ and $\hat{q}$ are $n \times n$ matrices.

The cases in which 
$[A_\mu,A_\nu] =c-number\equiv c_{\mu\nu}$
have a special meaning.
These correspond to BPS-saturated backgrounds \cite{WO}. Indeed, by setting
$\xi$ equal to $\pm \frac{1}{2}c_{\mu\nu} \Gamma^{\mu\nu}\epsilon$
in the $\cal{N}$=2 supersymmetry (\ref{Ssym1}) and (\ref{Ssym2}),
we obtain the relations
\beqa
(\delta^{(1)}\mp \delta^{(2)}) \psi &=&0 ,\n
(\delta^{(1)}\mp \delta^{(2)}) A_{\mu}&=& 0.
\eeqa
Namely, half of the supersymmetry is preserved in these backgrounds.
It is possible to construct higher dimensional solutions
which preserve half of the supersymmetry in an analogous way.
The D-branes in IIB matrix model have been investigated in
\cite{olesen}\cite{tseytlin}\cite{ishibashi}\cite{takata}.

The bosonic part of the action vanishes for the commuting matrices
$(A_{\mu})_{ij}=x^{\mu}_i \delta _{ij}$ where $i$ and $j$ are color indices.
These are the generic classical vacuum configurations of the model.
We have proposed to interpret $x^{\mu}_i$ as the space-time coordinates.
If such an interpretation is correct, the distributions of the eigenvalues
determine the extent and the dimensionality
of spacetime. Hence the structure of spacetime is dynamically determined by
the
theory.
As we have shown in \cite{AIKKT}, spacetime exits as a single bunch
and no single eigenvalue can escape from the rest.
However the appearance of a smooth manifold itself is not apparent
in this approach since we find four dimensional fractals in a simple
approximation.
Although it is very plausible that
gauge theory and gravitation may appear as low energy
effective theory, we are still not sure how matter fields propagate
\cite{ISK}.

The situation drastically simplifies if we consider noncommutative
backgrounds. These are the D-brane like solutions which preserve
a part of SUSY. Although the ultimate relevance of these solutions to
the vacuum of IIB matrix model is not clear, we can certainly test our
ideas to get a realistic model for space-time and matter with these
backgrounds.
We can indeed show that gauge theory appears as the low energy effective
theory.
In the case of $m$ coincident D-branes, we obtain noncommutative super-Yang
Mills
theory of 16 supercharges in the gauge group of $U(m)$.

It is of course well-known that the low energy effective action
for D-branes is super Yang-Mills theory.
If we mod out the theory with the translation operator,
we immediately find the corresponding super Yang-Mills theory
\cite{BFSS}\cite{WT}.
Noncommutative Yang-Mills theories have been obtained by the
compactification on noncummutative tori\cite{CDS}.
By `compactification', we may modify the theory by throwing
away many degrees of freedom. We are essentially left with gauge theory.
It is now well perceived through $AdS$/CFT correspondence that gauge theory
can represent gravitation in the vicinity of the D-branes\cite{maldacena}.
However gauge theory is not capable to describe gravitation in the flat
space-time far from the brane.

We point out that well-known twisted reduced models\cite{twisted}
are equivalent to noncommutative Yang-Mills theory.
The expansion around the infinitely extended D-branes in IIB matrix model
defines a twisted reduced model.
Using the equivalence,
we find noncommutative Yang-Mills theory in IIB matrix model.
We therefore propose that IIB matrix model with D-brane backgrounds
provides us a concrete definition of noncommutative Yang-Mills theory.
We point out that IIB matrix model contains
nonlocal degrees of freedom which can represent the gravitational
interaction in the flat ten dimensional space-time far from
the branes.
This fact will be demonstrated by calculating the potential between
a D-instanton and an anti-D-instanton on D3-branes.

The organization of this paper is as follows.
In section 2, we show that noncommutative Yang-Mills theory
is equivalent to large $N$ twisted reduced models.
In section 3, we apply the result of section 2 to IIB matrix  model
with D-brane backgrounds.
We study D-instantons in section 4. Section 5 is devoted
to conclusions and discussions. There we discuss the relation
between the Maldacena conjecture and the IIB matrix model conjecture.

\section{Noncommutative Yang-Mills as twisted reduced model}
\setcounter{equation}{0}

In this section, we show that noncommutative Yang-Mills theory
is equivalent to twisted reduced models.
Reduced models are defined by the dimensional reduction of $d$
dimensional gauge theory down to zero dimension (a point)\cite{RM}.
We consider $d$ dimensional $U(n)$ gauge theory coupled to
adjoint matter as an example:
\beq
S=-\int d^dx {1\over g^2}Tr({1\over 4}[D_{\mu},D_{\nu}][D_{\mu},D_{\nu}]
+{1\over 2}\bar{\psi}\Gamma _{\mu}[D_{\mu},\psi ]) ,
\eeq
where $\psi$ is a Majorana spinor field.
The corresponding reduced model is
\beq
S=- {1\over g^2}Tr({1\over4}[A_{\mu},A_{\nu}][A_{\mu},A_{\nu}]
+{1\over 2}\bar{\psi}\Gamma _{\mu}[A_{\mu},\psi ]) .
\eeq
Now $A_\nu$ and $\psi$ are $n\times n$ Hermitian matrices
and each component of $\psi$ is $d$-dimensional Majorana-spinor.
We expand the theory around the following classical solution.
\beq
[\hat{p}_{\mu},\hat{p}_{\nu}]=iB_{\mu\nu} ,
\eeq
where  $B_{\mu\nu}$ are $c$-numbers.
We remark that this commutation relation cannot be satisfied with finite
size matrices of dimension $n$. In fact it is spoiled at the boundary.
Nevertheless we can still assume it as long as we consider the degrees
of freedom which are localized in the region far from the boundary.
Although such a background is no longer a stable solution, its life
time becomes arbitrary large in the large $n$ limit.
We assume the rank of $B_{\mu\nu}$
to be $\tilde{d}$ and define its inverse $C^{\mu\nu}$ in $\tilde{d}$
dimensional subspace.
The directions orthogonal to the subspace
is called the transverse directions.
$\hat{p}_{\mu}$ satisfy the canonical commutation relations and
they span the $\tilde{d}$ dimensional phase space.
The semiclassical correspondence shows that the
volume of the phase space is $V_p=n(2\pi)^{\tilde{d}/2} \sqrt{detB}$.
Since we identify $\hat{p}_{\mu}$ as momenta, the phase space corresponds
to momentum space which is also called by the same name in particle physics.

We expand $A_{\mu}=\hat{p}_{\mu}+\hat{a}_{\mu}$. We Fourier decompose
$\hat{a}_{\mu}$ and $\hat{\psi}$ fields as
\beqa
\hat{a}&=&\sum_k \tilde{a}(k) exp(iC^{\mu\nu}k_{\mu}\hat{p}_{\nu}) ,\n
\hat{\psi}&=&\sum_k \tilde{\psi}(k) exp(iC^{\mu\nu}k_{\mu}\hat{p}_{\nu}) .
\label{twist}
\eeqa
The Hermiticity requires that $\tilde{a}^* (k)=\tilde{a}(-k)$ and
$\tilde{\psi} ^*(k)=\tilde{\psi} (-k)$.
Let us consider the case that $\hat{p}_{\mu}$ consist of $\tilde{d}/2$
canonical pairs $(\hat{p}_i,\hat{q}_i)$ which satisfy
$[\hat{p}_i,\hat{q}_j]=iB\delta_{ij}$.
We also assume that the  solutions possess the discrete symmetry
which exchanges canonical pairs and $(\hat{p}_i\leftrightarrow \hat{q}_i)$
in each canonical pair.
We then find $V_{p}=L^{\tilde{d}}$ where $L$ is the extension of each
$\hat{p}_{\mu}$.
The volume of the unit quantum in phase space is
$L^{\tilde{d}}/n=\lambda^{\tilde{d}}$
where $\lambda$ is the spacing of the quanta.
$B$ is related to $\lambda$ as $B=\lambda^2/(2\pi)$.
The eigenvalues of $\hat{p}_{\mu}$ are quantized in the unit of
$L/n^{2/ \tilde{d}}=\lambda/n^{1/ \tilde{d}}$.
So we restrict the range of $k_{\mu}$ as
$-n^{1/ \tilde{d}} \lambda/2<k_{\mu}< n^{1/ \tilde{d}}\lambda/2$.
Since $|\hat{p}_{\mu}| < L$, we can assume that $k_{\mu}$ is quantized in
the unit of $\lambda/n^{1/ \tilde{d}}$. So $\sum_k$ runs over $n^2$ degrees
of freedom
which coincide with those of $n$ dimensional Hermitian matrices.

We can construct a map from a matrix to a function
as
\beq
\hat{a} \rightarrow a(x)=\sum_k \tilde{a}(k) exp(ik_{\mu}x^{\mu}) .
\label{proj}
\eeq
We consider the product of two matrices
\beqa
&&\hat{a}=\sum_k \tilde{a}(k) exp(iC^{\mu\nu}k_{\mu}\hat{p}_{\nu}) ,\n
&&\hat{b}=\sum_k \tilde{b}(k) exp(iC^{\mu\nu}k_{\mu}\hat{p}_{\nu}) ,\n
&&
\hat{a}\hat{b}=\sum_{k,l} \tilde{a}(k)\tilde{b}(l) exp({1\over
2}C^{\mu\nu}k_{\mu}l_{\nu})
exp(iC^{\rho\sigma}(k_{\rho}+l_{\rho})\hat{p}_{\sigma}) .
\eeqa
By this construction, we obtain the $\star$ product
\beqa
\hat{a}\hat{b} &\rightarrow& a(x)\star b(x),\n
a(x)\star b(x)&\equiv&exp({iC^{\mu\nu}\over 2}{\partial ^2\over
\partial\xi^{\mu}
\partial\eta^{\nu}})
a(x+\xi )b(x+\eta )|_{\xi=\eta=0} .
\label{star}
\eeqa
The operation $Tr$ over matrices can be exactly mapped on the integration
over functions as
\beq
Tr[\hat{a}] =
\sqrt{det B}({1\over 2\pi})^{\tilde{d}\over 2}\int d^{\tilde{d}}x a(x) .
\label{traceint}
\eeq
It is easy to understand this statement if we consider the two dimensional
case.
\beqa
&&Tr(exp(iC^{\mu\nu}k_{\mu}\hat{p}_{\nu}))\n
&=&\int dq
<q|exp(ik_0\hat{p}/B)exp(-ik_1\hat{q}/B)exp(ik_0k_1/B)|q>\n
&=&2\pi B\delta (k_0)\delta (k_1) .
\eeqa
From these considerations, we find
the following map from matrices onto functions:
\beqa
\hat{a} &\rightarrow& a(x) ,\n
\hat{a}\hat{b}&\rightarrow& a(x)\star b(x) ,\n
Tr&\rightarrow&
\sqrt{det B}({1\over 2\pi})^{\tilde{d}\over 2}\int d^{\tilde{d}}x .
\label{momrule}
\eeqa

$P_{\mu}$ acts on $\hat{a}$ as
\beq
[\hat{p}_{\mu},\hat{a}]=\sum_k {k}_{\mu} \tilde{a}(k)
exp(iC^{\rho\sigma}k_{\rho}\hat{p}_{\sigma}) .
\label{pderiv}
\eeq
This motivates us to find the following correspondence:
\beq
[\hat{p}_{\mu}+\hat{a}_{\mu},\hat{o}]\rightarrow
{1\over i}\partial_{\mu}o(x)+a_{\mu}(x)\star o(x)-o(x)\star a_{\mu}(x) ,
\label{pcovder}
\eeq
The low energy excitations with $|k|<<\lambda$ are commutative
since
\beqa
&&[\hat{a}_{\mu},\hat{a}_{\nu}]\n
&=&
\sum_k \sum_l \tilde{a}(k)_{\mu}\tilde{a}(l)_{\nu}
[exp(iC^{\rho\sigma}k_{\rho}
\hat{p}_{\sigma})
,exp(iC^{\rho'\sigma'}l_{\rho'}\hat{p}_{\sigma'})]\n
&=&\sum_k \sum_l \tilde{a}(k)_{\mu}\tilde{a}(l)_{\nu}
exp(iC^{\rho\sigma}(k_{\rho}+l_{\rho})
\hat{p}_{\sigma})\n
&&\times
2i\sin({1\over 2}C^{\rho'\sigma'}k_{\rho'}l_{\sigma'}) .
\eeqa
We may interpret  the newly emerged coordinate space
as the semiclassical limit of $\hat{x}^{\mu}=C^{\mu\nu}\hat{p}_{\nu}$.
In such an interpretation
\beq
a(x)=Tr[\rho_x \hat{a}] ,
\label{semicl}
\eeq
where $\rho_x$ denotes a density matrix localized around the eigenvalue $x$.
Semiclassically we indeed find eqs. (\ref{proj}) and (\ref{traceint})
although we emphasize that eq.(\ref{momrule}) is the exact correspondence.

Applying the rule eq.(\ref{momrule}), the bosonic action becomes
\beqa
&&-{1\over 4g^2}Tr[A_{\mu},A_{\nu}][A_{\mu},A_{\nu}]\n
&=&
{\tilde{d}nB^2\over 4g^2}-\sqrt{det B}({1\over 2\pi})^{\tilde{d}\over 2}
\int d^{\tilde{d}}x
{1\over g^2} ({1\over 4}
[D_{\alpha},D_{\beta}][D_{\alpha},D_{\beta}]\n
&&+{1\over 2}[D_{\alpha},\varphi_{\nu}][D_{\alpha},\varphi_{\nu}]
+{1\over 4}[\varphi_{\nu},\varphi_{\rho}]
[\varphi_{\nu},\varphi_{\rho}])_{\star} .
\eeqa
In this expression, the indices $\alpha,\beta$ run over $\tilde{d}$
dimensional world volume directions  and $\nu,\rho$
over the transverse directions.
We have replaced $a_{\nu}\rightarrow\varphi_{\nu}$ in the transverse
directions. Inside $(~~)_{\star}$, the products should be understood as
$\star$
products and hence commutators do not vanish.
The fermionic action becomes
\beqa
&&{1\over g^2}Tr\bar{\psi}{\Gamma}_{\mu}[A_{\mu},\psi]\n
&=&
\sqrt{det B}({1\over 2\pi})^{\tilde{d}\over 2}
\int d^{\tilde{d}}x{1\over g^2}
(\bar{\psi}{\Gamma}_{\alpha}[D_{\alpha},\psi ]
+\bar{\psi}\Gamma_{\nu}[\varphi_{\nu},\psi ])_{\star} .
\eeqa
We therefore find noncommutative U(1) gauge theory.

In order to obtain noncommutative Yang-Mills theory with $U(m)$ gauge group,
we consider new classical solutions which are obtained by replacing each
element
of $\hat{p}_{\mu}$ by the $m\times m$ unit matrix:
\beq
\hat{p}_{\mu} \rightarrow \hat{p}_{\mu}\otimes \mbox{\boldmath $1$}_m .
\eeq
The fluctuations around this background $\hat{a}$ and $\hat{\psi}$
can be Fourier
decomposed in the analogous way as in eq.(\ref{twist}) with  $m$ dimensional
matrices $\tilde{a}(k)$ and $\tilde{\psi} (k)$ which satisfy
$\tilde{a}(-k)=\tilde{a}^{\dagger}(k)$ and
$\tilde{\psi}(-k)=\tilde{\psi} ^{\dagger}(k)$.
It is then clear that $[\hat{p}_{\mu}+\hat{a}_{\mu},\hat{o}]$ can be mapped
onto the
nonabelian covariant derivative $D_{\mu}o(x)$ once we use $\star$ product.
Applying our rule (\ref{momrule}) to the action in this case,
we obtain
\beqa
&&{\tilde{d}nB^2\over 4g^2}-\sqrt{det B}({1\over 2\pi})^{\tilde{d}\over 2}
\int d^{\tilde{d}}x
{1\over g^2} tr({1\over 4}
[D_{\alpha},D_{\beta}][D_{\alpha},D_{\beta}]\n
&&+{1\over 2}[D_{\alpha},\varphi_{\nu}][D_{\alpha},\varphi_{\nu}]
+{1\over 4}[\varphi_{\nu},\varphi_{\rho}][\varphi_{\nu},\varphi_{\rho}]\n
&&+{1\over 2}\bar{\psi}{\Gamma}_{\alpha}[D_{\alpha},\psi ]
+{1\over 2}\bar{\psi}\Gamma_{\nu}[\varphi_{\nu},\psi ])_{\star} .
\eeqa
The Yang-Mills coupling is found to be $(2\pi )^{\tilde{d}\over
2}g^2/B^{\tilde{d}/2}$.
Therefore it will decrease if the density of quanta in phase space
decreases with fixed $g^2$.

Although our  arguments here has been in the continuum theory,
it is straightforward to generalize our arguments to lattice gauge
theory by replacing
\beq
exp(iC^{\mu\nu}k^{min}_{\mu}\hat{p}_{\nu})
\rightarrow U_{\mu} ,
\eeq
where the index $\mu$ should not be summed.
$U_{\mu}$ are t'Hooft matrices whose canonical pairs satisfy
\beq
U_{\mu}U_{\nu}=U_{\nu}U_{\mu}exp({2\pi i\over n^{2/\tilde{d}}}) ,
\eeq
since $|k^{\min}_{\mu}|=\lambda /n^{1/\tilde{d}}$ as we have explained
in this section.
Here we need to remark on the novelty of our interpretation of twisted
reduced models since it has been interpreted as large $N$ limit
of $U(N)$ gauge theory. Our innovation is that we have constructed the
coordinate space in the matrices by the relation
$\hat{x}^{\mu}=C^{\mu\nu}\hat{p}_{\nu}$.
The remarkable feature of our construction is the appearance of the
coordinate space out of momentum space. It is thanks to the noncommutativity
of momentum space. We may interpret that the noncommutativity effectively
introduces the maximum momentum scale $\lambda$.
Since $\hat{x}^{\mu}=C^{\mu\nu}\hat{p}_{\nu}$, the large momentum
region corresponds to large length scale in the dual $\hat{x}^{\mu}$ space.
Thus the physics beyond this momentum scale is better understood
in the dual coordinate space. Therefore the whole
construction reminds us string theory and T duality.
In fact we will see  in the following section that the identical structure
emerges in association with the D-branes in IIB matrix model.

\section{Noncommutative Yang-Mills and D-branes}
\setcounter{equation}{0}

In this section we apply the results of the preceding section
to D-branes.
We notice that the D-string solution in eq.(\ref{Sstaticsolution})
is precisely the type of the backgrounds in twisted reduced models
we have considered in the preceding section.
As we have found in section 2, the both momentum space and coordinate
space are embedded in the matices of twisted reduced models.
They are related by $\hat{x}^{\mu}=C^{\mu\nu}\hat{p}_{\nu}$.
This relation should be understood in the follwing sense.
The plane wave with a wave vector $k_{\mu}$ corresponds to
an eigenstate of $\hat{P}_{\mu}=[\hat{p}_{\mu},~]$
with $k_\mu$ as the eigenvalue.
They are commutative to each other since $|k_\mu| < \lambda$.
The dual coordinate space is
embedded in the large eigenvalues of the matrices which are rotated by
$C^{\mu\nu}$. They are also commutative.
We obtain the same physics if we interpret $k_{\mu}$ as
momenta or $C^{\mu\nu}\hat{p}_{\nu}$ as coordinates $\hat{x}^{\mu}$.

We need to interpret $A_{\mu}$ as coordinates in IIB matrix model
due to $\cal{N}$=2 SUSY as we have emphasized in the introduction.
For this purpose, we identify the solution of IIB matrix model as
$\hat{x}^{\mu}$ which satisfy.
\beq
[\hat{x}^{\mu},\hat{x}^{\nu}]=-iC^{\mu\nu} .
\eeq
Now the plane waves correspond to
the eigenstates of $\hat{P}_{\mu}$ with small eigenvalus, where
$\hat{p}_{\mu}=B_{\mu\nu}\hat{x}^{\nu}$.
$\hat{x}^{\mu}$ and $\hat{p}_{\nu}$ satisfy the canonical commutation
relation: $[\hat{x}^{\mu},\hat{p}_{\nu}]=i\delta^{\mu}_{\nu}$.
We expand $A_{\mu}=\hat{x}_{\mu}+\hat{a}_{\mu}$ as before and
$\hat{a}_{\mu}$ and
$\hat{\psi}$ can be Fourier decomposed as in
\beqa
\hat{a}&=&\sum_k \tilde{a}(k) exp(i{k_{\mu}}\hat{x}^{\mu}),\n
\hat{\psi}&=&\sum_k \tilde{\psi}(k) exp(i{k_{\mu}}\hat{x}^{\mu}).
\label{twistbr}
\eeqa

The space-time translation is realized by the following Unitary
operator:
\beqa
&&exp(i\hat{p}_{\nu}d^{\nu})(\hat{x}^{\mu}+\hat{a}^{\mu})
exp(-i\hat{p}_{\nu}
d^{\nu})\n
&=&\hat{x}^{\mu}+d^{\mu}+
\sum_k \tilde{a}(k) exp(i{k_{\nu}}d^{\nu})exp(i{k_{\rho}}\hat{x}^{\rho}) .
\eeqa
We find that
$\tilde{a}(k)$ is multiplied by the phase $exp(i{k_{\nu}}d^{\nu})
\tilde{a}(k)$.

Once the eigenvalues of $\hat{P}_{\mu}$ are identified with
$k_{\mu}$,
the coordinate space has to be embedded in the rotated matrices
as we have seen in section 2.
If we identify the large eigenvalues of $A^{\mu}$ as the coordinates
$x^{\mu}$,
we have to rotate the covariant derivatives as follows:
\beq
[\hat{x}^{\mu}+\hat{a}^{\mu},\hat{o}]
\rightarrow C^{\mu\nu}({1\over i}\partial_{\nu}o(x)+b_{\nu}(x)\star o(x)
-o(x)\star b_{\nu}(x)).
\label{covder}
\eeq
Note that we have defined a new gauge field $b_{\mu}(x)$ by this expression.
We can map the matrices onto functions by using
the rule eq.(\ref{momrule}) which is derived in the preceding
section.

We consider the gauge invariance on D-string.
The IIB matrix model is invariant under the Unitary transformation:
$A_{\mu}\rightarrow UA_{\mu}U^{\dagger},
\psi \rightarrow U\psi U^{\dagger}$. As we shall see, the gauge
symmetry can be embedded in the $U(n)$ symmetry.
We expand $U=exp(i\hat{\lambda} )$ and parameterize
\beq
\hat{\lambda}=\sum_k \tilde{\lambda} (k) exp(i{k_0}\hat{x}^0+i{k_1}\hat{x}^1) .
\eeq
Under the gauge transformation, we find the fluctuations around
the fixed background transform as
\beqa
\hat{a}^{\mu}&\rightarrow & \hat{a}^{\mu}+i[\hat{x}^{\mu},\hat{\lambda} ]
-i[\hat{a}^{\mu},\hat{\lambda} ]\n
&=&\sum_k exp(i{k_0}\hat{x}^0+i{k_1}\hat{x}^1)
(\tilde{a}(k)^{\mu}+iC^{\mu\nu}{k}_{\nu}\tilde{\lambda} (k))\n
&+&\sum_k\sum_l \tilde{a}(k)^{\mu}\tilde{\lambda} (l)
exp(i{(k_0+l_0)}\hat{x}^0+i{(k_1+l_1)}\hat{x}^1)
{i(k_0l_1-k_1l_0)\over B}+\cdots .
\eeqa
\beqa
\hat{\psi}&\rightarrow & \hat{\psi} -i[\hat{\psi},\hat{\lambda} ]\n
&=&\sum_k exp(i{k_0}\hat{x}^0+i{k_1}\hat{x}^1)
\tilde{\psi} (k)\n
&+&\sum_k\sum_l \tilde{\psi} (k)\tilde{\lambda} (l)
exp(i{(k_0+l_0)}\hat{x}^0+i{(k_1+l_1)}\hat{x}^1)
{i(k_0l_1-k_1l_0)\over B}+\cdots .
\eeqa
We interpret the above result as
\beqa
&&b_{\alpha}(x) \rightarrow b_{\alpha}(x) +
{\partial \over \partial x^{\alpha}}\theta (x)
+\eta_{\beta}{\partial b_{\alpha}(x)\over \partial x^{\beta}}
+\cdots ,\n
&&a_{\nu}(x) \rightarrow a_{\nu}(x)+
\eta_{\beta}{\partial a_{\nu}(x)\over \partial x^{\beta}}
+\cdots ,\n
&&\psi (x)\rightarrow \psi (x)+
\eta_{\beta}{\partial \psi (x)\over \partial x^{\beta}}+\cdots ,
\eeqa
where $\eta_{\alpha}=C^{\alpha\beta}
\partial _{\beta}\lambda (x)$.
We indeed find $U(1)$ gauge group in the commutative limit.
The leading corrections in $1/B$ represent the volume preserving
diffeomorphism.

Applying the rule eq.(\ref{momrule}), the bosonic action becomes
\beqa
&&-{1\over 4g^2}Tr[A_{\mu},A_{\nu}][A_{\mu},A_{\nu}]\n
&=&
{LT\over B 4\pi g^2}-
{1\over B^2 2\pi g^2}\int d^2x ({1\over 4B}
[D_{\alpha},D_{\beta}][D_{\alpha},D_{\beta}]\n
&&+{1\over 2}[D_{\alpha},\varphi_{\nu}][D_{\alpha},\varphi_{\nu}]
+{B\over 4}[\varphi_{\nu},\varphi_{\rho}]
[\varphi_{\nu},\varphi_{\rho}])_{\star} ,
\eeqa
where $D_{\alpha}={\partial/(i\partial x^{\alpha})}
+b_{\alpha}$ and $a_{\nu}=\sqrt{1/B}\varphi_{\nu}$.
The fermionic action becomes
\beqa
&&Tr\bar{\psi}{\Gamma}_{\mu}[A_{\mu},\psi]\n
&=&
\int d^2x tr(\bar{\psi}\tilde{\Gamma}_{\alpha}[D_{\alpha},\psi ]
+{\sqrt{B}}\bar{\psi}\Gamma_{\nu}[\varphi_{\nu},\psi ])_{\star} ,
\eeqa
where ${\Gamma}_{\alpha}=\epsilon_{\alpha\beta}\tilde{\Gamma}_{\beta}$.
We therefore find noncommutative two dimensional $\cal{N}$=8
$U(1)$ gauge theory.

We next consider $m$ parallel D-strings.
We can construct such a solution $A^{cl}_{\mu}$ by replacing
each element of a D-string solution by an $m$ by $m$ unit matrix.
As before we decompose $A_{\mu}=A^{cl}_{\mu}+\hat{a}_{\mu}$ where each element
of $\hat{a}_{\mu}$ is now an $m$ by $m$ matrix.
We can simply apply the generalized rule which implies
$[A_{\alpha},~]\rightarrow C^{\alpha\beta}D_{\beta}$ where
$D_{\alpha}$ is the nonabelian covariant derivative.
The bosonic action becomes
\beqa
&&-{1\over 4g^2}Tr[A_{\mu},A_{\nu}][A_{\mu},A_{\nu}]\n
&=&
{mLT\over B4\pi g^2}-
{1\over B^22\pi g^2}\int d^2x tr({1\over 4B}
[D_{\alpha},D_{\beta}][D_{\alpha},D_{\beta}]\n
&&+{1\over 2}[D_{\alpha},\varphi_{\nu}][D_{\alpha},\varphi_{\nu}]
+{B\over 4}[\varphi_{\nu},\varphi_{\rho}]
[\varphi_{\nu},\varphi_{\rho}])_{\star} .
\eeqa
The symbol $tr$ implies taking the trace over $U(m)$ gauge group.
The fermionic action becomes
\beqa
&&Tr\bar{\psi}{\Gamma}_{\mu}[A_{\mu},\psi]\n
&=&
\int d^2x tr(\bar{\psi}\tilde{\Gamma}_{\alpha}[D_{\alpha},\psi ]
+{\sqrt{B}}\bar{\psi}\Gamma_{\nu}[\varphi_{\nu},\psi ])_{\star} .
\eeqa
We therefore find two dimensional $\cal{N}$=8 super Yang-Mills theory
with $U(m)$ gauge group.

We move on to consider higher dimensional D-branes.
A D3-brane may be constructed as follows:
\beqa
A_0 &=& \frac{T}{\sqrt{2\pi n_1}}\hat{q} \equiv \hat{x}^0 ,\n
A_1 &=& \frac{L}{\sqrt{2\pi n_1}}\hat{p} \equiv \hat{x}^1 ,\n
A_2 &=& \frac{L}{\sqrt{2\pi n_2}}\hat{q}' \equiv \hat{x}^2 ,\n
A_3 &=& \frac{L}{\sqrt{2\pi n_2}}\hat{p}' \equiv \hat{x}^3 ,\n
\mbox{other }A_{\mu} \mbox{'s}&=& 0,
\eeqa
which may be embedded into $n_1n_2$ dimensional matrices.
We can further consider $m$ parallel D3-branes after  replacing each
element
of the D3-brane solution by $m$ by $m$ unit matrix.
Under the replacements $A_{\alpha}\rightarrow C^{\alpha\beta}
D_{\beta}$ and
$A_{\nu}\rightarrow (1/B)\varphi_{\nu}$,
the bosonic action becomes
\beqa
&&-{1\over 4g^2}Tr[A_{\mu},A_{\nu}][A_{\mu},A_{\nu}]\n
&=&{mTL^3\over (2\pi )^2 g^2}
-{1\over B^2(2\pi )^2g^2}\int d^4x tr({1\over 4}[D_{\alpha},D_{\beta}]
[D_{\alpha},D_{\beta}]\n
&&+{1\over 2}[D_{\alpha},\varphi_{\nu}][D_{\alpha},\varphi_{\nu}]
+{1\over 4}[\varphi_{\nu},\varphi_{\rho}]
[\varphi_{\nu},\varphi_{\rho}])_{\star} ,
\eeqa
where we integrate over the four dimensional world volume of
D3-branes.
As for the fermionic action, we find
\beqa
&&Tr\bar{\psi}\Gamma_{\mu}[A_{\mu},\psi]\n
&=&
\int d^4x tr(\bar{\psi}\tilde{\Gamma}_{\alpha}[D_{\alpha},\psi ]
+\bar{\psi}\Gamma_{\nu}[\varphi_{\nu},\psi ])_{\star} .
\eeqa
We thus find four dimensional $\cal{N}$=4 super Yang-Mills theory.
The Yang-Mills coupling is found to be
$g^2B^2$.
Recall that $(2\pi /B)^2 =R^4$ is the unit volume of a quantum
and $R$ is the average spacing.
If we consider a background with
larger $R$ for fixed $g^2$, we find that the Yang-Mills coupling decreases.

In this section, we have obtained noncommutative Yang-Mills theory
with D-brane backgrounds in IIB matrix model
to all orders in power series of $1/B$.
Alternatively we can view the IIB matrix model with D-brane backgrounds
as a concrete definition of noncommutative Yang-Mills theory.

\section{D-instantons in IIB matrix model}
\setcounter{equation}{0}

As we have shown in section 3,
the low energy effective theory with D3-backgrounds are
super Yang-Mills theory. In super Yang-Mills theory, there are local
nontrivial solutions, namely instantons.
The equation of motion in IIB matrix model is
\beq
[A_{\mu},[A_{\mu},A_{\nu}]]=0 .
\eeq
With our substitution rule, $A_{\alpha}\rightarrow
C^{\alpha\beta}D_{\beta}$, we obtain,
\beq
[D_{\alpha},[D_{\alpha},D_{\beta}]]=0 .
\eeq
Since the instantons are nontrivial solutions of
gauge theory, they must become those of IIB matrix model
after ramifications at short distances.
As we have shown, short distance modification of Yang-Mills theory
in IIB matrix model is to render it noncommutative.
In fact such solutions on noncommutative $R^4$ are constructed in \cite{NS}.

The 't Hooft solution is
\beqa
A_{\mu}(x)&=&i\Sigma_{\mu\nu}\Phi (x)^{-1}\partial_{\nu}\Phi (x) ,\n
\Phi (x) &=&1+{\rho^2\over |x-x_i|^2} ,
\label{instanton}
\eeqa
where $\Sigma_{\mu\nu}$ is self-dual in $\mu\nu$ and takes values in
traceless two by two Hermitian matrices.
The location of the instanton is denoted by $x_i$ and $\rho$ is its size.
The prescription is just replace $\Phi (x)$ by its noncommutative analog.
Although it is an interesting problem to study small instantons whose size
is comparable to $R$, it suffices to consider instantons whose size is much
larger
than $R$ in this section.


Let us consider an instanton solution first.
The classical value of the action is
\beq
S={2TL^3 \over g^2(2\pi )^2}+{1\over B^2g^2} .
\label{d3inst}
\eeq
We interpret the first and the second term of eq.(\ref{d3inst})
as the action of two D3-branes and that of a D-instanton
respectively.
It preserves one fourth of the supersymmetry and hence
eq.(\ref{d3inst}) receives no quantum corrections.
In order to see whether the solution preserves a part of supersymmetry,
we consider
\beqa
\delta^{(1)}\psi &=& \frac{i}{2}
                     [A_{\mu},A_{\nu}]\Gamma_{\mu\nu}\epsilon ,\n
&=& \frac{i}{2}(-iC^{\alpha\beta}+
C^{\alpha\gamma}C^{\beta\delta}[D_{\gamma},D_{\delta}]
{(1 \mp \Gamma^5 )\over 2})\Gamma_{\alpha\beta}
\epsilon  ,\n
\delta^{(2)}\psi &=& \xi
\eeqa
where $\mp$ correspond to self-dual and anti self-dual field strengths
respectively.
Therefore an (anti)instanton solution preserves one fourth of the
supersymmetry which satisfy
\beqa
\Gamma^5 \epsilon  &=& \pm \epsilon  ,\n
\xi &=& \frac{1}{2}C^{\alpha\beta}\Gamma_{\alpha\beta}\epsilon  .
\eeqa
This argument is valid for generic (anti)self-dual field configurations.

We next consider a classical solution of IIB matrix model
which represents an instanton and an (anti)instanton.
We can realize $U(4)$ gauge theory by considering four D3 branes.
We embed an instanton into the first $SU(2)$ part and the other
(anti)instanton into the remaining $SU(2)$ part. We separate
them in the fifth dimension by the distance $b$:
\beqa
A_0&=&\left( \begin{array}{cc}
\hat{x}^0+a_0 & 0 \\
0 & \hat{x}^0+a'_0
\end{array}
\right) ,\n
A_1&=&\left( \begin{array}{cc}
\hat{x}^1+a_1 & 0 \\
0 & \hat{x}^1+a'_1
\end{array}
\right) ,\n
A_2&=&\left( \begin{array}{cc}
\hat{x}^2+a_2 & 0 \\
0 & \hat{x}^2+a'_2
\end{array}
\right) ,\n
A_3&=&\left( \begin{array}{cc}
\hat{x}^3+a_3 & 0 \\
0 & \hat{x}^3+a'_3
\end{array}
\right) ,\n
A_4&=&\left( \begin{array}{cc}
{b\over 2} & 0 \\
0 & -{b\over 2}
\end{array}
\right) ,\n
A_{\rho}&=&0 ,
\eeqa
where $\rho=5,\cdots ,9$.

The classical action is twice of eq.(\ref{d3inst}).
While two instanton system receives no quantum corrections,
the instanton - anti-instanton system
receives quantum corrections since it is no longer BPS.
We now evaluate the one loop effective potential due to an instanton
and (anti)instanton.
Since they are local excitations, they must couple to gravity.
These solutions are characterized by the adjoint field strength
$F_{\mu\nu}$
which does not vanish at the locations of the instantons.
Let us assume that they are separated by a long distance compared to their
sizes. We also assume that $b >> R$.
Then we can choose two disjoint blocks
in each of which a large part of an (anti)instanton is contained.
Let the location and the size of instantons $(x_i,\rho_i)$
and  $(x_j,\rho_j )$.
The ten dimensional distance of them is $r^2=(x_i-x_j)^2+b^2$.
Here we have assumed that $r >> \rho$.
The potential between the $i$-th and the $j$-th blocks
due to the off-diagonal $(i,j)$ block has been calculated
in \cite{IKKT}.
\beqa
W^{(i,j)}&=&
{1\over r^8}
(-Tr^{(i,j)}(F_{\mu\nu}F_{\nu\lambda}F_{\lambda\rho}F_{\rho\mu})
-2Tr^{(i,j)}(F_{\mu\nu}F_{\lambda\rho}F_{\mu\rho}F_{\lambda\nu})\n
&&+{1\over 2}Tr^{(i,j)}(F_{\mu\nu}F_{\mu\nu}F_{\lambda\rho}F_{\lambda\rho})
+{1\over 4}Tr^{(i,j)}(F_{\mu\nu}F_{\lambda\rho}F_{\mu\nu}F_{\lambda\rho})
\n
&=&{3\over 2 r^8}
( -n_j \tilde{b}_8(f^{(i)})- n_i \tilde{b}_8(f^{(j)}) \n
&&
-8Tr(f_{\mu\nu}^{(i)}f_{\nu\sigma}^{(i)})
Tr(f_{\mu\rho}^{(j)}f_{\rho\sigma}^{(j)})
+Tr(f_{\mu\nu}^{(i)}f_{\mu\nu}^{(i)})
Tr(f_{\rho\sigma}^{(j)}f_{\rho\sigma}^{(j)})\n
&&+Tr(f_{\mu\nu}^{(i)}\tilde{f}_{\mu\nu}^{(i)})
Tr(f_{\rho\sigma}^{(j)}\tilde{f}_{\rho\sigma}^{(j)})) ,
\label{blockin}
\eeqa
where $\tilde{f}_{\mu\nu}=\epsilon_{\mu\nu\rho\sigma}f_{\rho\sigma}/2$
and
\beq
\tilde{b}_8(f)=
{2\over 3}(Tr(f_{\mu\nu}f_{\nu\lambda}f_{\lambda\rho}f_{\rho\mu})
+2Tr(f_{\mu\nu}f_{\lambda\rho}f_{\mu\rho}f_{\lambda\nu})
-{1\over 2}Tr(f_{\mu\nu}f_{\mu\nu}f_{\lambda\rho}f_{\lambda\rho})
-{1\over 4}Tr(f_{\mu\nu}f_{\lambda\rho}f_{\mu\nu}f_{\lambda\rho})) .
\eeq
The novel feature of
eq.(\ref{blockin}) is that we have kept axion type interactions also.
Note that the $\tilde{b}_8(f)=0$ for an (anti)instanton configuration.
So the potential between an instanton and an (anti)instanton is
\beq
{3\over 2 r^8}
(-8Tr(f_{\mu\nu}^{(i)}f_{\nu\sigma}^{(i)})
Tr(f_{\mu\rho}^{(j)}f_{\rho\sigma}^{(j)})
+Tr(f_{\mu\nu}^{(i)}f_{\mu\nu}^{(i)})
Tr(f_{\rho\sigma}^{(j)}f_{\rho\sigma}^{(j)})
+Tr(f_{\mu\nu}^{(i)}\tilde{f}_{\mu\nu}^{(i)})
Tr(f_{\rho\sigma}^{(j)}\tilde{f}_{\rho\sigma}^{(j)})
) .
\label{instint}
\eeq
Here we can apply the low energy approximation such as
\beqa
&&Tr(f_{\mu\nu}^{(i)}f_{\mu\nu}^{(i)})=
{1\over B^2(2\pi)^2}\int d^4xtr([D^i_{\mu},D^i_{\nu}][D^i_{\mu},D^i_{\nu}])
=-{R^4\over \pi^2} ,\n
&&Tr(f_{\mu\nu}^{(i)}\tilde{f}_{\mu\nu}^{(i)})=
{1\over B^2(2\pi)^2}\int d^4xtr\epsilon_{\mu\nu\rho\sigma}
([D^i_{\mu},D^i_{\nu}][D^i_{\rho},D^i_{\sigma}])
=\mp{R^4\over \pi^2} ,\n
&&Tr(f_{\mu\nu}^{(i)}f_{\nu\rho}^{(i)})=
{1\over B^2(2\pi)^2}\int d^4xtr([D^i_{\mu},D^i_{\nu}]
[D^i_{\nu},D^i_{\rho}])=-{R^4\over 4\pi^2}\delta_{\mu\rho} ,
\eeqa
where $D^i_{\mu}$ denotes the covariant derivative
of the instanton background which is localized at the $i$-th block.
So the interactions eq.(\ref{instint})
can be interpreted due to the exchange of
dilaton, axion and gravitons.
We find that the potential between two instantons vanish due to
their BPS nature.
On the other hand,
the following potential is found between an instanton and an anti-instanton
\beq
-{3\over \pi ^4}{R^8\over r^8} .
\label{gravin}
\eeq

We remark that the above approximation is no longer valid
when $b<R$. In this case the interactions between an instanton and
an anti-instanton is well described by the gauge fields which are low
energy
modes of IIB matrix model. They are close to diagonal degrees of freedom
in IIB matrix model.
Their contribution can be estimated by gauge theory.
On the other hand when $b>>R$, the standard gauge theory description
breaks down since we have to take account of the noncommutativity.
In that case, we believe that the block-block interaction gives us a
correct result. It is in a sense T dual description to gauge theory.
The one loop effective potential can be calculated by gauge theory when
$b<<R$
\beq
\Gamma = -{1\over B^42(4\pi )^2b^4}\int d^4x b_8 ,
\eeq
where we have assumed $b\rho >> 1/B$.
The above expression is estimated
as follows:
\beqa
\Gamma &=&-{144\over B^4\pi^2b^4}\int d^4x {\rho^4\over ((x-y)^2+\rho^2)^4}
{\rho^4\over ((x-z)^2+\rho^2)^4}\n
&&\sim -{24\rho^4\over  B^4r^{8} b^{4}} ,
\label{gaugein}
\eeqa
where $r=|y-z|$ is assumed to be much larger than $\rho$.
We note that eq.(\ref{gaugein}) falls off with the identical power for
large $r$  with eq.(\ref{gravin}).

In $AdS$/CFT correspondence, the instanton size is interpreted as the
radial coordinate of a D-instanton in $AdS_5$\cite{BGKR}.
With this interpretation, the D-instanton approaches
the boundary of $AdS_5$ when the instanton size vanishes.
The gravitational interaction between D-instantons in flat space
is known to be of the form $\alpha '^4/r^8$.
If we assume that $\alpha ' \sim 1/B$ and $g^2 \sim g_s \alpha '^2$,
we find that eq.(\ref{gravin}) indeed of such a type.
In $AdS_5$, it is modified as
\beq
{\alpha '^4\over L^8}{\rho ^4\rho '^4\over ((y-z)^2+(\rho-\rho ')^2)^4} ,
\eeq
where $L$ is the radius of $AdS_5$.
Here we notice a limitation of gauge theory.
We cannot describe gravitational interaction between a
D-instanton and an anti-D-instanton
in flat space far from the branes by gauge theory.
On the other hand the IIB matrix model can describe the gravitational
interaction between them in both regions
near and far from the branes.
Hence we find an important advantage of IIB matrix model over gauge theory
as a nonperturbative formulation of superstring.
It is due to the existence of nonlocal (or noncommutative) degrees of
freedom  in IIB matrix model.

We have proposed that the fundamental strings are created by the
Wilson loop operators. Simple examples are the following vertex operators
for a dilaton, an axion and gravitons which are consistent with
eq.(\ref{instint}):
\beqa
&&Tr\{[A_{\alpha},A_{\beta}][A_{\alpha},A_{\beta}]
exp(ik_{\gamma} A_{\gamma})\}
+ \mbox{fermionic terms} ,\n
&&\epsilon_{\alpha\beta\gamma\delta}
Tr\{[A_{\alpha},A_{\beta}][A_{\gamma},A_{\delta}]
exp(ik_{\gamma'} A_{\gamma'})\}
+ \mbox{fermionic terms} ,\n
&&Tr\{[A_{\alpha},A_{\mu}][A_{\mu},A_{\beta}]
exp(ik_{\gamma} A_{\gamma})\}
+ \mbox{fermionic terms} .
\label{highver}
\eeqa
We have the corresponding vertex operators in CFT:
\beqa
&&\int d^4xtr([D_{\alpha},D_{\beta}][D_{\alpha},D_{\beta}])exp(ik\cdot x)
+ \mbox{fermionic terms},\n
&&\int d^4x\epsilon_{\alpha\beta\gamma\delta}
tr([D_{\alpha},D_{\beta}][D_{\gamma},D_{\delta}])exp(ik\cdot x)
+ \mbox{fermionic terms},\n
&&\int d^4xtr([D_{\alpha},D_{\beta}][D_{\beta},D_{\gamma}])exp(ik\cdot x)
+ \mbox{fermionic terms} .
\label{dagop}
\eeqa

Under the replacements $A_{\alpha}\rightarrow C^{\alpha\beta}
D_{\beta},A_{\nu}\rightarrow 1/B\varphi_{\nu}$,
the Wilson loops in IIB matrix model coincide with those
in CFT such as
\beqa
&&Tr\{[A_{\alpha},A_{\mu}][A_{\mu},A_{\beta}]
exp(ik_{\gamma} (\hat{x}_{\gamma}+\hat{a}_{\gamma}))\}\n
&\rightarrow&
\int d^4 x tr\{[D_{\alpha},D_{\gamma}][D_{\gamma},D_{\beta}]+
[D_{\alpha},\varphi_{\nu}][D_{\beta},\varphi_{\nu}]\}exp(ik\cdot x)
+\cdots ,
\label{lowver}
\eeqa
where we have replaced $\hat{x}\rightarrow x$ in the argument
of the exponential.
It is interesting to investigate in more detail
the above noticed relationship between the  Wilson loops in
IIB matrix model and the vertex operators in CFT.



\section{Conclusions and discussions}
\setcounter{equation}{0}
In this paper we have shown that large $N$ twisted reduced models
can be interpreted in terms of noncommutative Yang-Mills theory.
Such a system appears in IIB matrix model as
infinitely extended D-brane solutions which preserve a part of SUSY.
We therefore obtain noncommutative Yang-Mills theory with such
backgrounds. We propose that IIB matrix model with such a
background provides us a precise definition of noncommutative
Yang-Mills theory.
The novel feature is that the real coordinate space and the conjugate
momentum space can be embedded into matrices in the large $n$ limit.
We have studied D-instanton interactions.
When they overlap, their interaction is well described by gauge
theory. When they are well separated in the fifth dimension, the gauge
theory description breaks down. In that situation, IIB matrix model
provides an accurate description and we find the result is consistent
with IIB supergravity. In this sense, IIB matrix model is valid in both
the gauge theory region and supergravity regions. Therefore we need not
assume the overlap of the two unlike $AdS$/CFT correspondence.

Let us consider the implication of our results on
Maldacena conjecture\cite{maldacena}.
We have shown that $U(N)$ gauge theory is obtained if we consider
$N$ parallel D-branes in IIB matrix model whose
distances are much smaller compared to string scale.
The basic conjecture of IIB matrix model is that it is a
nonperturbative formulation of IIB superstring theory.
Its low energy limit is IIB supergravity.
The tree level string theory is considered to be obtained by
summing planar diagrams and string perturbation theory
is identified with the topological expansion of the matrix model.
If we consider the large $N$ limit with fixed t'Hooft coupling,
we are left with the planar diagrams.
Therefore the Maldacena conjecture follows from our IIB matrix
model conjecture.
We can also point out the limitations of Maldacena conjecture.
The gauge theory description is valid only when the distances
between the D-branes are much smaller than the string scale.
Therefore it is not applicable when the branes are separated by
the distance much longer than the string scale.
In other words, CFT cannot describe the flat space far from the brane.
In this regard, IIB matrix model has the definite advantage as we have
demonstrated by the instanton interactions.

Although we have obtained 4 dimensional gauge theory
with D-brane backgrounds, we find ten dimensional gravity.
It is presumably not a bad news for our scenario to get realistic
models out of IIB matrix model, since D3-brane backgrounds possess
enormous vacuum energy density $(1/\alpha ')^2$.
This is certainly what has been expected from string perturbation theory.
Although the eigenvalue distributions of $A_{\mu}$ are
sharply peaked at four dimensional hyperplane, they presumably spread
out into the transverse directions due to the massless scalars.
It suggests us to consider the solutions without massless scalars
(broken SUSY?) to obtain four dimensional gravity. In any case,
noncommutative geometry may be a crucial element to obtain gauge theory
in IIB matrix model.

We conclude this section with the following comment.
Chepelev and Tseytlin argued that our D-brane solutions can
be interpreted as not pure D-branes but those with nonvanishing $U(1)$
field strength\cite{tseytlin}.
In this paper we have shown that our solution can be interpreted in terms
of noncommutative geometry with vanishing $U(1)$ field strength.
Recently Seiberg and Witten has pointed out that noncommutative Yang-Mills
is equivalent to ordinary Yang-Mills theory with a
nonvanishing $B$-field or $U(1)$ field strength\cite{string99}.
Apparently these arguments are consistent with each other
and reflect the T duality of string theory.

\begin{center} \begin{large}
Acknowledgments
\end{large} \end{center}
The final part of this work was carried out while we got together at
PIMS of UBC.
We would like to thank the organizers of the conference
PFS '99:T.Lee, G. Semenoff and especially Y. Makeenko
for their warm hospitality.
This work is supported in part by the Grant-in-Aid for Scientific
Research from the Ministry of Education, Science and Culture of Japan.

\end{document}